\documentclass{article}
\usepackage[utf8]{inputenc}
\usepackage{authblk}

\usepackage{relsize}
\usepackage{xspace}
\usepackage[inline]{enumitem}

\usepackage{amsmath,amsfonts}

\usepackage{url}
\usepackage{xcolor}

\newtheorem{example}{Example}[section]

\newcommand\mopac{MOPaC\xspace}

\title{MOPaC: The Multiple Offers Protocol for Multilateral Negotiations with Partial Consensus}

\author{Pradeep K. Murukannaiah and Catholijn M. Jonker}
\affil{Interactive Intelligence, TU Delft, The Netherlands}
\affil{\{p.k.murukannaiah, c.m.jonker\}@tudelft.nl}

\date{}

\begin{document}

\maketitle

\begin{abstract}
    Existing protocols for multilateral negotiation require a full consensus among the negotiating parties. In contrast, we propose a protocol for multilateral negotiation that allows partial consensus, wherein only a subset of the negotiating parties can reach an agreement. We motivate problems that require such a protocol and describe the protocol formally.
\end{abstract}

\section{Introduction}

Multilateral negotiations enable group decisions involving two or more parties. The SAOP and AMOP protocols \cite{Aydogan-2017-ACAN-NegoProtocols} facilitate multilateral negotiations, but require a (full) consensus among parties. However, negotiations requiring full consensus can be too restrictive, e.g., in the following scenarios.

\begin{example}
\label{ex:meeting}
Consider a meeting scheduling negotiation among $n$ ($>2$) parties. Ideally, we would like all parties to agree on the meeting but the meeting can take place even if a few parties cannot make it. That is, if there are no offers that all parties agree on, but there are offers that a (large) subset of the parties agree on, the negotiation can still succeed with a partial consensus.
\end{example}

\begin{example}
\label{ex:government}
Consider the process of coalition building in a multi-party political system \cite{Dupont-2996-IntNego-NegotiationasCoalitionBuilding}. In an election, $n$ ($>2$) parties contested, and a certain number of candidates won from each party. However, none of the parties have a majority, on their own. Thus, the parties must negotiate and build a coalition to form the government. A partial consensus is sufficient to build a successful coalition.
\end{example}

\begin{example}
\label{ex:flatmate}
Consider that a group of $n$ ($>2$) friends would like to share (rent) flats with each other during their visit to a city. Each person in the group has preferences on the characteristics of the flat he or she wants to rent and the characteristics of their flatmates. The friends engage in a negotiation to find suitable flatmates. This negotiation can yield more than one successful deal each involving a partial consensus among a subset of the friends.
\end{example}

\section{The {\mopac} Protocol}


Let $A$ be a set of $n$ agents in a multilateral negotiation. $\forall i, A_i: 0 < i \leq n$, $A_i$, identifies one of the agents, where $\forall i, j: 0 < i,j \leq n$ and $i \neq j \implies A_i \neq A_j$.

Each agent $A_i$ has a power $p_i \in \mathbb{N}$ in the negotiation. Thus, the power, $p$, of a full consensus among agents is:
\begin{equation}
    p_{max} = \sum_{i=1}^{n} p_i
\end{equation}

Let $p_{min}$ be the minimum power required for any (partial) consensus to form in this negotiation, which is sent as a parameter to the protocol.

\subsection{Protocol}
A multilateral negotiation with the \mopac protocol runs for one or more rounds. Each round consists of the following phases.

\begin{description}[leftmargin=1em]
\item [Bidding phase:] Let $B$ be a bidding space of $m$ possible bids. Each agent puts a bid on the table. Let $b_{ij} \in B$ be agent $A_i$'s bid in round $j$.

After the bidding phase, each agent is communicated the bid and power of each agent. That is, the following list is communicated to each agent:
\begin{equation}
 \Big[ [ A_1, b_{1j}, p_1],~\ldots [ A_n, b_{nj}, p_n] \Big]
\end{equation}

\item [Voting phase:] Each agent votes (\textsl{accept}/\textsl{reject}) on each bid on the table. With each \textsl{accept} vote, an agent must indicate:
\begin{enumerate}[label=(\arabic*),noitemsep]
    \item a \emph{minimum consensus threshold}, $C_{min}$: $p_{min} \le C_{min} \le p_{max} \in \mathbb{N}$, and
    \item  a \emph{maximum consensus threshold}, $C_{max}$, $C_{min} \le C_{max} \le p_{max} \in \mathbb{N}$.
\end{enumerate}

That is, the two possible votes an agent can provide on a bid $b_i$ (including the agent's own bid) are:
\begin{equation}
    v_i =
    \begin{cases}
      \langle b_i, \mathrm{accept}, C_{min}, C_{max} \rangle, \text{or} \\
      \langle b_i, \mathrm{reject} \rangle.
    \end{cases}
\end{equation}

By accepting a bid $b_i$ with $C_{min} = m$ and $C_{max} = n$, an agent indicates that it will accept the bid $b_i$ if a partial consensus group of \emph{power} at least $m$ and at most $n$ forms around bid $b_i$ (Section~\ref{sec:partial-consensus}).

After the voting phase,  each agent is communicated the votes of each agent. That is, the following list is communicated to each agent:
\begin{equation}
 \Big[ [ A_1, \langle b_1, v_1 \rangle,~\ldots \langle b_m, v_m \rangle],~\ldots [ A_n, \langle b_1, v_1 \rangle,~\ldots \langle b_m, v_m \rangle] \Big]
\end{equation}

\item [Opt-in phase:] Each agent votes again, as before, but with constraints that an agent cannot \textsl{reject} or reduce the $C_{min}$ value of a bid it had \textsl{accept}ed in the voting phase. That is, if an agent accepted $b_i$ with vote $v_i$ in the voting phase, its vote for $b_i$ in the opt-in phase $v'_i$ should be as follows.
\begin{align}
\forall v_i: v_i &= \langle b_i, \mathrm{accept}, C_{min}, C_{max} \rangle,\\ \nonumber
v'_i &= \langle b_i, \mathrm{accept}, C_{min} \le C'_{min} \le p_{max}, C_{min} \le C'_{max} \le p_{max} \rangle.
\end{align}
However, an agent can \textsl{accept} a bid it had \textsl{reject}ed in the voting phase.
\begin{align}
\forall v_i: v_i &= \langle b_i, \mathrm{reject} \rangle,\\ \nonumber
v'_i &=
\begin{cases}
      \langle b_i, \mathrm{reject} \rangle, \text{or} \\
      \langle b_i, \mathrm{accept}, p_{min} \le C'_{min} \le p_{max}, C'_{min} \le C'_{max} \le p_{max} \rangle.
\end{cases}
\end{align}


\item [Continuation or termination:] There is more than one way a \mopac negotiation can terminate. The following are two potential options. In both of these options, the continuation or termination decision is made after computing the viable consensus groups at the end of the second round of voting.

\begin{enumerate}[label=(\arabic*)]
    \item Determine a viable consensus group with the largest power (see Section~\ref{sec:partial-consensus}). If there is a tie for the group with largest power, break the ties randomly. For the agents in the chosen viable consensus group, the negotiation terminates with a deal. For the remaining agents, the negotiation terminates without a deal. This kind of termination is ideal, e.g., when negotiating to schedule a meeting (Example~\ref{ex:meeting}) or form a government (Example~\ref{ex:government}).
    \item Determine a viable consensus group with the largest power (see Section~\ref{sec:partial-consensus}). If there is a tie for the group with largest power, break the ties randomly. Determine if there are more viable groups consisting of the remaining agents. If so, the negotiation terminates with a deal for such agents, too. The remaining agents go to the next round of negotiation. The negotiation continues until the deadline or until one or no agent remains in this round, whichever happens first. This type of termination is ideal, e.g., when negotiation for flatmates (Example~\ref{ex:flatmate}).
\end{enumerate}

\end{description}

\section{Partial Consensus Formation}
\label{sec:partial-consensus}

Given a set of agents $C = \{ A_1^C, \ldots A_m^C \}$, who each voted \textsl{accept} for a bid $b_i$, $C$ is a partial consensus group on $b_i$. The \textbf{power} of the consensus $C$ is:

\begin{equation}
    p_C = \sum_{A_i \in C} p_i.
\end{equation}

\noindent A consensus $C$ for a bid $b_i$ is \textbf{viable} iff $\forall A_i \in C$, $C_{min} \le p_C \le C_{max}$. That is, a viable consensus group for a bid $b_i$ consists of a set of agents that accepted $b_i$ and the power of the group is within the minimum and maximum consensus thresholds indicated by each agent in the group.

\subsection{Computing Viable Consensus Groups}
The following is a naive approach for computing all viable consensus groups at the end of a MOPaC round. Although naive, this approach should be computationally feasible for a small number (e.g., $n \le 10$) of agents.

\begin{enumerate}
    \item Enumerate all bids $b_i$ in a round.
    \item Compute the set of all possible agent groups $C_P^A$ such that a group of agents $c_i \in C^A$ iff $c_i \in P^A$, the power set of $A$ and $|c_i| > 1$.

    $|C^A| = 2^n - n - 1$.

    \item For each bid $b_i$ and agent group $c_i$, determine if $c_i$ is (1) a consensus group, and (2) viable. If so, add $\langle b_i, c_i \rangle$ to the list of viable groups, and proceed.
\end{enumerate}

\textbf{Optimization note}: The apriori algorithm used in association analysis \cite{Tan-2006-DataMining} can be used to prune the search space. That is, if a set $c_i$ is not a consensus group, none of its subsets will form a consensus group.


 \providecommand{\url}[1]{#1}

\end{document}